\documentclass[prd]{revtex4}
 
\usepackage{graphicx}
\usepackage{dcolumn}
\usepackage{bm}

\def \be{\begin{equation}}
\def \ee{\end{equation}}
\def \bea{\begin{eqnarray}}
\def \eea{\end{eqnarray}}
 
\begin{document}
 
\title{Numerical stability of a new conformal-traceless 3+1 formulation\\
of the Einstein equation}
 
\author{Pablo Laguna and Deirdre Shoemaker}
\affiliation{Center for Gravitational Physics and Geometry, and\\
Center for Gravitational Wave Physics\\
Penn State University, University Park, PA 16802}
 
\begin{abstract}
There is strong evidence indicating that the particular form used to recast the
Einstein equation as a 3+1 set of evolution equations has a fundamental impact
on the stability properties of numerical evolutions involving black holes
and/or neutron stars. Presently, the longest lived 
evolutions have been obtained using
a parametrized hyperbolic system developed by Kidder, Scheel and Teukolsky or
a conformal-traceless system introduced by Baumgarte, Shapiro,
Shibata and Nakamura. We present 
a new conformal-traceless system. 
While this new system has some elements in
common with the Baumgarte-Shapiro-Shibata-Nakamura system, 
it differs in both the type of conformal transformations and
how the non-linear terms involving the extrinsic curvature are handled. 
We show results from 3D numerical evolutions of
a single, non-rotating black hole in which 
we demonstrate that this new
system yields a significant improvement
in the life-time of the simulations.
\end{abstract}
 
 
\maketitle

\section{Introduction}

So far, every 3D effort to perform 3+1
numerical simulations of space-times containing black-hole 
singularities, or compact objects such 
as neutron stars, has encountered the frustration of producing 
short-lived evolutions. The impediment is due to the presence of
sever instabilities. There is a common agreement among researchers
that the origin of these instabilities
is not purely numerical. 
Inconsistent boundary conditions far away
from the holes or
neutron stars, unsuitable choices of gauge or coordinate conditions, 
inappropriate algorithms to excise the black hole singularity, etc.
are all contributing factors.  
Since approximately the mid 90's, the numerical relativity community
has been accumulating evidence supporting the view that a
crucial aspect determining the life-time of simulations is
the form of the Einstein equation that is implemented in the numerical codes.

When viewed as an initial value problem, the Einstein equation can be cast
in what is commonly known as a 3+1 form; that is, the Einstein equation becomes 
a set of constraint and evolution equations that govern 
the history of the geometry of space-like
hypersurfaces, geometrodynamics. There are many ways of writing 
the Einstein equation in a 3+1 form. 
During the early years of numerical relativity,
the most popular form was that introduced
by Arnowitt, Deser and Misner (ADM) \cite{ADM}.
It is safe to say that all of the attempts to obtain 
long-term stable and convergent,
3D simulations of
black holes using codes based on the ADM formulation have failed.
These efforts have produced short lived evolutions lasting $< 200M$, with
$M$ the mass of the black hole. Unfortunately, the reasons behind
this failure are still not known.  Nonetheless, 
because of all of these failed attempts,
the numerical community has practically abandoned the ADM system. 

One alternative to ADM has been formulations of the Einstein equation
that are explicitly hyperbolic \cite{hyper_rev}. 
These formulations 
have been introduced in part to facilitate 
handling both the outer boundary of the computational domain and the
excision boundary in the vicinity of the black hole singularity. 
Hyperbolic formulations certainly have the advantage that, by knowing
the characteristics of each of the evolved fields, one is able
to apply the correct boundary conditions to those fields 
that require them. Hyperbolic formulations also enjoy the advantage  
of the mathematical machinery to prove existence, uniqueness and
well-posedness. Unfortunately, 
hyperbolic formulations have so far been able to produce
evolutions of single black holes 
lasting $\sim 600M\,-\,1300M$ \cite{KST}.  

Another formulation of the Einstein equation that has been relatively 
successful is the one re-introduced by Baumgarte and Shapiro \cite{BS} and
originally developed by Shibata and Nakamura \cite{SN}. This formulation,
commonly known as the BSSN system, belongs to
the class of conformal traceless formulations (see next section
for a detailed derivation). There are several ingredients that go 
into the derivation of these equations. 
Besides the use of conformal transformations and a 
traceless decomposition of the extrinsic curvature, 
perhaps the most important aspect is the introduction of
a connection. With this connection, the formulation contains 
some hyperbolic flavor.
Studies have shown, however, that 3D, single black hole evolutions are
still limited to life-times $\sim 500M$ \cite{miguel}.

Because of the relative success of the hyperbolic and BSSN systems
over the ADM formulation, there seems to be a certain degree of consensus that 
a successful formulation should have some hyperbolic
flavor. The formulation we introduce in this paper retains this property.
The innovative aspect of this system enters 
in shifting the focus from the 
principal part of the evolutions equations, i.e. those terms containing 
differentiation, to non-linear algebraic terms.  
Our work is motivated by a recent numerical study
of evolutions of single black holes in spherical symmetry
\cite{Kelly} using the ADM system. In Ref.~\cite{Kelly}, long-term 
stable and convergent evolutions were achieved by ``adjusting'' the ADM system.
The adjusting consisted of eliminating or switching the sign of
non-linear terms involving
the extrinsic curvature. The stability that was obtained was remarkable.
Unfortunately, an extension to 3D
is not possible because the adjustment took advantage of the
spherical symmetry in the problem. As we will show, it is possible to 
achieve a similar elimination or switching of sign of non-linear terms.  
Instead of working with the ADM system, one needs to consider
a modified form of the BSSN formulation.
We will present 3D numerical simulations of
single, non-rotating  black holes in which we
demonstrate a remarkable improvement in the
stability and duration of the evolutions when compared with
those based on BSSN.

\section{New Conformal Traceless Equations}

In order to gain a better understanding of the elements present in 
the conformal traceless system here considered, we will review the
derivation of the BSSN formulation.
The traditional derivation of the BSSN system starts with the
ADM formulation. The ADM primary variables are the
spatial metric, $g_{ij}$, and extrinsic curvature, $K_{ij}$, of the
space-like hypersurfaces. The equations that govern the dynamics of
$g_{ij}$ and $K_{ij}$ are 
\bea
\partial_o g_{ij}&=& - 2\, \alpha\, K_{ij} 
\label{gdot} \\
\partial_o K_{ij}&=&  - \nabla_i
\nabla_j \alpha +   \alpha\,R_{ij}\nonumber \\
& +& \alpha\,(K\, K_{ij} - 2 K_{ik}\, K^k\,_j) \,,\label{Kdot}
\eea
where
$
\partial_o \equiv \partial_t -{\cal L}_{\beta} 
$
and $K=g_{ij}K^{ij}.$
Although we concentrate our attention on the vacuum case,
the inclusion of matter sources is straightforward.
Above, $\nabla_i$ and $R_{ij}$ denote covariant differentiation
and Ricci tensor associated with $g_{ij}$, respectively. The
lapse function, $\alpha$, and shift vector, $\beta^i$,
encapsulate the diffeomorphism freedom of General Relativity 
and are thus freely specifiable. In terms of the ADM variables,
the Einstein constraints read
\bea
R + K^2 - K_{ij} K^{ij} &=& 0  \label{hamcons}\\
\nabla_j K^{ij} - \nabla^i K & =& 0 \, .\label{momcons}
\eea
As customary in numerical relativity,
evolutions are performed without ensuring that the evolved data
continue to satisfy the constraints. Eqs. (\ref{hamcons}) and (\ref{momcons})
are only used to monitor the accuracy and stability of the simulations.

The first step in obtaining the BSSN formulation 
is to abandon $g_{ij}$ and $K_{ij}$ as
primary variables and work instead with $\Phi,\, \hat g_{ij},\,
K$ and $\hat A_{ij}$. The relationships between these and the ADM variables are
\bea
\label{eq:Phi}
\Phi         &=& \frac{1}{6}\,\ln g^{1/2} \\
\label{eq:ghat}
\hat g_{ij}  &=& e^{- 4 \Phi}\, g_{ij} \\
\label{eq:A}
\hat A_{ij}  &=& e^{- 4 \Phi}\,A_{ij}\,,
\eea
where
$A_{ij}  = K_{ij} - g_{ij}\,K/3$.
The first point to notice is that (\ref{eq:Phi}) and (\ref{eq:ghat}) imply that
the conformal metric $\hat g_{ij}$ has unit determinant. Enforcing this condition
in numerical evolutions has an important impact on the stability of the simulations. 
Unless noted, indices will be raised and lowered with the conformal
metric $\hat g_{ij}$.
Given (\ref{eq:Phi})-(\ref{eq:A}), it is not difficult to show that the
ADM system (\ref{gdot})-(\ref{Kdot}) of evolution equations
take the form
\bea
\label{Phidot}
\partial_o \Phi &=& - \frac{1}{6} \alpha\, K \\
\label{gdot2}
\partial_o \hat g_{ij} &=& - 2\, \alpha\, \hat A_{ij} \\
\label{Kdot2}
\partial_o K &=& - \nabla_i \nabla^i \alpha +
        \alpha\,(\hat A_{ij} \hat A^{ij} + K^2/3) \\
\label{Adot}
\partial_o \hat A_{ij} & = & e^{- 4 \Phi} \left(
        - \nabla_i \nabla_j \alpha  +
        \alpha\, R_{ij} \right)^{TF} \nonumber\\
        &+& \alpha\, (K \hat A_{ij} - 2 \hat A_{il} \hat A^l_{~j})\,.
\eea

Before continuing, there are several points to be made about the derivation 
and structure of these equations. First, in deriving Eq.~(\ref{Kdot2}), 
the Hamiltonian constraint~(\ref{hamcons}) was used to
eliminate the Ricci scalar.  
In Eq.~(\ref{Adot}), the superscript $TF$ denotes the trace-free part
of the tensor between brackets, e.g. 
$T_{ij}^{TF} \equiv T_{ij} - g_{ij} g^{kl}T_{kl}/3$
for any tensor $T_{ij}$. When calculating $R_{ij}^{TF}$, 
the trace $R$ is again eliminated using the Hamiltonian
constraint~(\ref{hamcons}). To simplify notation, we did not explicitly 
transform the terms involving 
covariant derivatives of the lapse or 
the Ricci tensor, $R_{ij}$. These terms should be understood as
being given respectively by:
\bea
\label{lapse}
\nabla_i \nabla_j \alpha &=&  \widehat\nabla_i \widehat\nabla_j \alpha 
- 4\widehat\nabla_{(i}\Phi\widehat\nabla_{j)}\alpha \nonumber\\
&+& 2\, \hat g_{ij} \widehat\nabla^l\Phi\widehat\nabla_l\alpha \\
\label{ricci}
R_{ij} &=& \hat R_{ij} 
-2 \widehat \nabla_i \widehat \nabla_j \Phi -
2 \hat g_{ij} \widehat \nabla^l \widehat \nabla_l \Phi \nonumber\\
&+& 4 \widehat \nabla_i \Phi\widehat \nabla_j \Phi
        - 4\, \hat g_{ij} \widehat \nabla^l \Phi \widehat \nabla_l \Phi\,,
\eea
where $\hat R_{ij}$ is the Ricci tensor computed from $\hat g_{ij}$.
One should also keep in mind that $\hat g_{ij}$ and $\hat A_{ij}$ 
are tensor densities of weight $-2/3$, so their Lie derivatives are
\be
{\cal L}_{\beta}\, \hat A_{ij} = \beta^k \partial_k \hat A_{ij}
        + \hat A_{ik} \partial_j \beta^k
        + \hat A_{kj} \partial_i \beta^k
        - \frac{2}{3} \hat A_{ij} \partial_k \beta^k 
\ee
and a similar expression for $\hat g_{ij}$. 
The function $\Phi$ is not a true scalar,
see Eq.~(\ref{eq:Phi}). 
Its Lie derivative is given by
\be
{\cal L}_{\beta}\, \Phi  = \beta^k \partial_k \Phi + \frac{1}{6}
\,\Phi\partial_k \beta^k\,.   
\ee

Next is perhaps the key ingredient of the BSSN system. One introduces a 
conformal connection
\be \label{cgsf}
\widehat \Gamma^i \equiv \hat g^{jk} \widehat \Gamma^{i}_{jk}
	= - \partial_j \hat g^{ij},
\ee
where the $\widehat \Gamma^{i}_{jk}$ are the connection coefficients
associated with $\hat g_{ij}$. The last equality
holds because of the condition $\hat g = 1$. 
The motivation for introducing these connections is to substitute in 
the Ricci tensor $\hat R_{ij}$ all the derivatives of 
$\hat g^{jk} \widehat \Gamma^{i}_{jk}$ in favor of derivatives
of $\widehat \Gamma^i$. In doing so, the Ricci tensor yields
a system of evolution equations \cite{BS,SN} with ``hyperbolic flavor.''

Expanding the number of primary variables in the system
to include the conformal
connections implies the need of an additional evolution equation.
From definition (\ref{cgsf}), one can
show that the evolution equation for $\widehat\Gamma^i$ is given by
\bea
\label{eq:Gamma}
\partial_o \widehat\Gamma^i &=& \hat g^{jk}\partial_{jk}\beta^i
+ \frac{1}{3}\hat g^{ij}\partial_{jk}\beta^k
- 2\,\hat A^{ij}\partial_j\alpha \nonumber\\
&+& 2\,\alpha\widehat\Gamma^i_{jk}\hat A^{jk}
+ 12\,\alpha\hat A^{ij}\partial_j\Phi 
- \frac{4}{3}\alpha\widehat\nabla^iK\,.
\eea
In deriving this equation, the momentum constraint (\ref{momcons})
was used to eliminate derivatives of $\hat A^{ij}$. A final point to be made, 
$\widehat\Gamma^i$ is a vector density of weight 2/3, so
\be
{\cal L}_{\beta}\, \widehat\Gamma^i  = \beta^k \partial_k \widehat\Gamma^i 
- \widehat\Gamma^k\partial_k \beta^i + \frac{2}{3}
\,\widehat\Gamma^i\partial_k \beta^k\,.   
\ee
In general terms, the BSSN system involves: conformal transformations, 
a trace-free
decomposition of the extrinsic curvature, the introduction of 
a conformal connection and the use of constraints to eliminate the
Ricci scalar and derivatives of the trace-free extrinsic curvature. 

We are now prepared to derive the new system of evolution equations. 
Because this system retains the main ingredients of the BSSN formulation,
its derivation proceeds along a
similar path. However, the starting point is modified.
Instead of (\ref{eq:Phi})-(\ref{eq:A}), we use 
\bea
\label{nct:Phi}
\Phi         &=& \frac{1}{6}\,\ln g^{1/2} \\
\label{nct:ghat}
\hat g_{ij}  &=& e^{- 4 \Phi}\, g_{ij} \\
\label{nct:K}
\hat K       &=& e^{6\,n\,\Phi}K \\
\label{nct:A}
\hat A^i\,_j   &=& e^{6\,n\,\Phi}A^i\,_j \\
\label{nct:lapse}
N            &=& e^{-6\,n\,\Phi}\alpha \,,
\eea
with $n$ a parameter to be determined later.
Notice the main differences. We allow for the trace of the extrinsic curvature and 
the lapse function to be conformally transformed. $\hat K$ and $N$ are densities. Also, 
we use as primary variable the mixed-index form of the
trace-less part of the extrinsic curvature, $\hat A^i\,_j$.
If $n=0$, one recovers the BSSN scalings.
With the above transformations, the evolution equations take the following form
\bea
\label{nct:Phidot}
\partial_o \Phi &=& - \frac{1}{6} N\, \hat K \\
\label{nct:gdot2}
\partial_o \hat g_{ij} &=& - 2\, N\, \hat A_{ij} \\
\label{nct:Kdot2}
\partial_o \hat K &=& -e^{6\,n\,\Phi} \nabla_i \nabla^i \alpha 
+ N\,\hat A^i\,_j \hat A^j\,_i\nonumber\\ 
&+& (1-3\,n)\,N\,\hat K^2/3 \\
\label{nct:Adot}
\partial_o \hat A^i\,_j & = & e^{6\,n\, \Phi} \left(
- \nabla^i \nabla_j \alpha  +
\alpha\, R^i\,_j \right)^{TF}\nonumber\\
&+& (1-n)\,N\,\hat K \hat A^i\,_j \\
\label{nct:Gamma}
\partial_o \widehat\Gamma^i &=& \hat g^{jk}\partial_{jk}\beta^i
+ \frac{1}{3}\hat g^{ij}\partial_{jk}\beta^k
- 2\,\hat A^{ij} \partial_j N \nonumber \\
&+& 2\,N\widehat\Gamma^i_{jk}\hat A^{jk}
- \frac{4}{3}\,N\,\hat g^{ij}\partial_j\hat K \nonumber\\
&+& 12\,(1-n)\,N\hat A^{ij}\partial_j\Phi \,.
\eea
Here again, the terms involving $\alpha$ and $R_{ij}$ should be understood as
being given by (\ref{lapse}) and (\ref{ricci}), with the index raised with
$g^{ij}$. In addition, the lapse expression (\ref{lapse})
should be modified with the appropriate substitution
in terms of the densitized lapse $N$.

By fixing the attention to Eq.~(\ref{nct:Adot}), one can understand the
motivation behind the particular form of the conformal scalings and the use
of $\hat A^i\,_j$ as a primary variable. 
If we choose $n\ge 1$, the term $N\,\hat K \hat A^i\,_j$ in Eq.~(\ref{nct:Adot})
could be eliminated or have its sign switched.
In the spherically symmetric numerical study in Ref.~\cite{Kelly}, 
it was demonstrated that this algebraic term was 
responsible for the instabilities of the ADM-based simulations.
In Ref.~\cite{Kelly}, the changes in sign or eliminations of these terms 
were obtained by substitution of the 
momentum constraint.
This adjustment of the ADM system is not possible in non-spherical symmetry.
However, it is clear that the conformal 
scalings we have introduced accomplish the same effect.

The importance of eliminating $N\,\hat K \hat A^i\,_j$ can
be understood by the following hand-waving argument. 
Equation (\ref{nct:Adot}) has the following structure
\bea
\label{eq:hand}
\partial_t \hat A^i\,_j &=& C\, \hat A^i\,_j\nonumber\\
&+& \partial\hat A\,\hbox{Terms} +\hbox{Terms without}\,\hat A
\eea
where
\be
C \equiv n\,\partial_k\beta^k + (1-n)\,N\,\hat K\,.
\ee
At the continuum level, for a single, static black hole, the
r.h.s. of Eq. (\ref{eq:hand}) should vanish identically. However, because
of truncation errors, this will not be the case in numerical evolutions.
If this cancellation does not take place, a solution of the
form $\hat A^i\,_j \propto e^{C\,t}$ is in principle allowed.
Thus, $C>0$ has the potential of yielding an exponential growth. 
Almost all the numerical evolutions of single black holes have been performed
with space-time foliations such that $K>0$. Since the simulations of physical
interested are future directed (i.e. $\alpha > 0$), then $ N\,\hat K > 0$. 
As a consequence, the second term in $C$ has the ``wrong'' sign. Because of the
conformal scalings we have introduced, values of
$n \ge 1$ correct this sign problem. 
The first term in $C$ gives the impression that it also has the wrong sign.
This is not the case. All of the single black hole solutions of interest
have shift vectors for which $\partial_k\beta^k < 0$.
For more generic, multiple black hole situations, it should be possible to
construct shift vectors that preserve this property. If not, a large enough
value of $n$ should be able to yield the appropriate sign of $C$.
Finally, we point out that this hand-waving
argument provides a reasonable explanation of why
the simulations in \cite{Kconst} with $K = \hbox{const} < 0$ were
significantly more stable than those with positive $K$.

The use of a mixed-index, traceless extrinsic curvature in 3D studies is not new. 
One of us used this
approach in numerical studies of inhomogeneous inflationary space-times
\cite{KLM}. The evolution equations used in \cite{KLM} did not include a conformal
connection $\widehat \Gamma^i$ nor a densitized lapse. The use of densitized 
variables had a different motivation in \cite{KLM}.
With those densitized variables, the shift terms in the evolution equations could be
written in an almost flux-conservative form. 
However, it was quickly realized that, because during
an inflationary epoch the determinant of $g_{ij}$ grows exponentially, working
with positive weight, densitized variables becomes eventually problematic.
Our attempts to obtain stable 3D, single black hole evolutions with
a code based on the evolution system in Ref.~\cite{KLM} have failed.
Given the success of the results in the present work, we believe that the failure
of the system in \cite{KLM} is likely due to the absence of the
conformal connection $\widehat \Gamma^i$.  
There are indications that in spherical symmetry, using flux-conservative equations 
seems to extend the life of the simulations \cite{Luis}. This view is also supported
by recent work using a hyperbolic formulation \cite{David} in which the
evolution equations are written in a flux-conservative form.
Finally, mixed-index extrinsic curvature, not traceless, as a primary variable
has been used in 1D evolutions of plane-symmetric cosmologies \cite{Centrella}
and gravitational waves \cite{Bardeen}, 
supporting the view that this form is more advantageous. 
 
\section{Stability Experiments}

To test the stability of the new system, we concentrate
our attention on a single, non-rotating black hole. 
We use the ingoing-Eddington-Finkelstein form of the black hole solution.
The black hole singularity is handled via excision. That is, 
a region inside the horizon, containing the black hole singularity, is
removed from the computational domain.  Our excision has a cubical shape.
At its boundary, the values
of the primary variables $( \Phi,\, \hat g_{ij},\, \hat K,\, \hat A^i\,_j
,\,\widehat\Gamma^i)$ are obtained by third order extrapolation of the
corresponding evolved values. 
At the outer boundary, the evolved data are blended to the analytic solution.
Temporal updating is performed via the iterated Crank-Nicholson method.
After each time-step, we impose the constraints 
$\hat g = 1$ and $\hat A^i\,_i = 0$ as follows:
\bea
\hat A^i\,_j &\rightarrow& \hat A^i\,_j - \hat g^i\,_j \hat A^k\,_k/3 \\ 
\hat g_{ij} &\rightarrow& \hat g^{-1/3}\,\hat g_{ij}\,.  
\eea
Once the new $\hat g_{ij}$ is obtained, we use definition 
(\ref{cgsf}) to also reset $\widehat\Gamma^i$.

We do not write the shift terms in an almost flux-conservative form
as done in Ref.~\cite{KLM}. All the discretization of spatial operators
is done with centered finite differences. The exception is 
shift terms of the form $\beta^k\partial_k$. Those terms are approximated 
using an up-wind discretization described in Ref.~\cite{Kelly}. 

Since it is well known that the choice of gauge or coordinate conditions
has a fundamental impact on the duration of the evolutions, we decided to
test the new system on the most challenging case, namely the gauge condition
in which $N$ and $\beta^i$ are obtained from the
exact analytic solution.  

\begin{figure*}[fig1]
\includegraphics{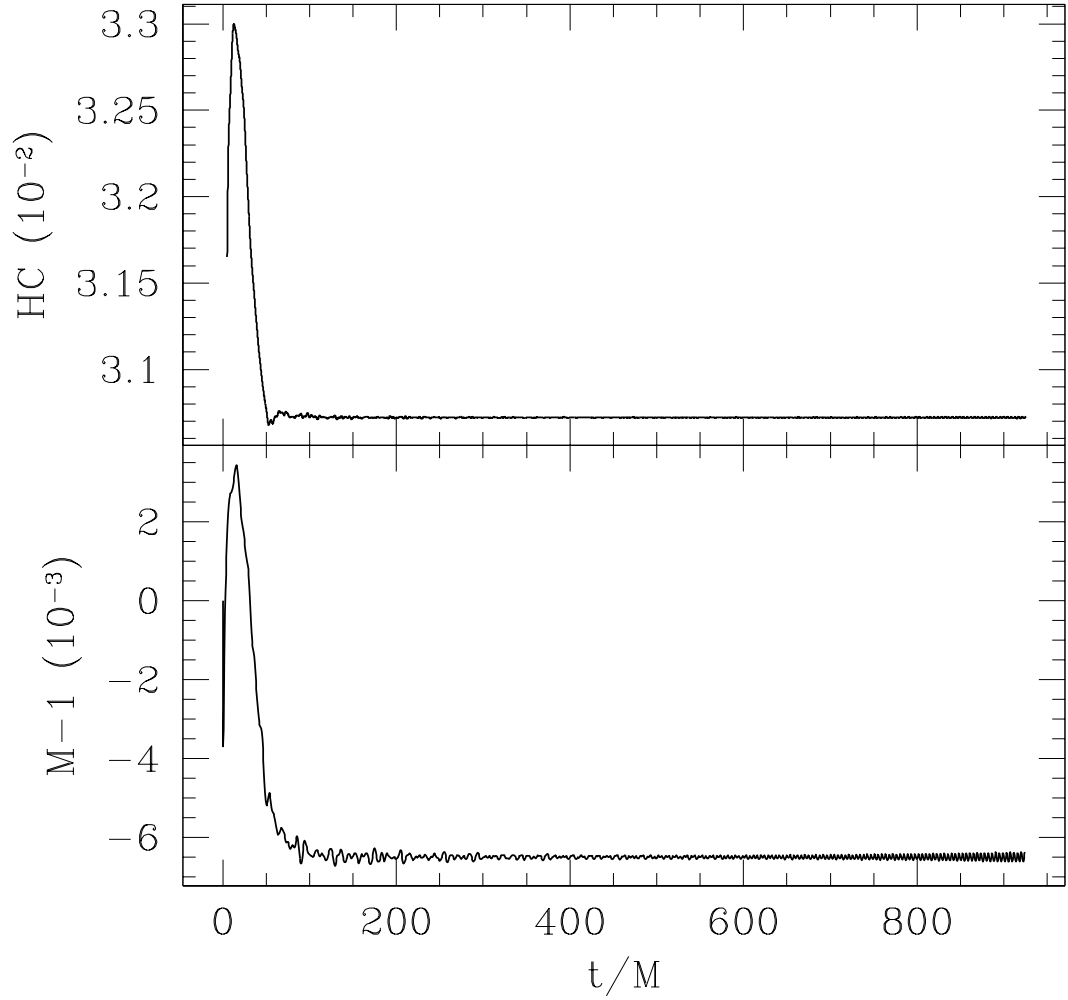}
\caption[figure1]{\label{fig1}
L2 norm as a function of time 
of the Hamiltonian constraint (HC) and the error in the mass of the
black hole ($M-1$) from a 3D evolution 
of a single, non-rotating black hole in ingoing-Eddington-Finkelstein coordinates. 
The outer boundary of the 
computational domain was located at $9M$ with $M$ the mass of the black hole
in the initial data. The grid-point spacing in this run was 
$0.25M$. The run was performed imposing reflection symmetries across
the $x=0$ and $y=0$ planes.
}\end{figure*}

\begin{figure*}[fig2]
\includegraphics{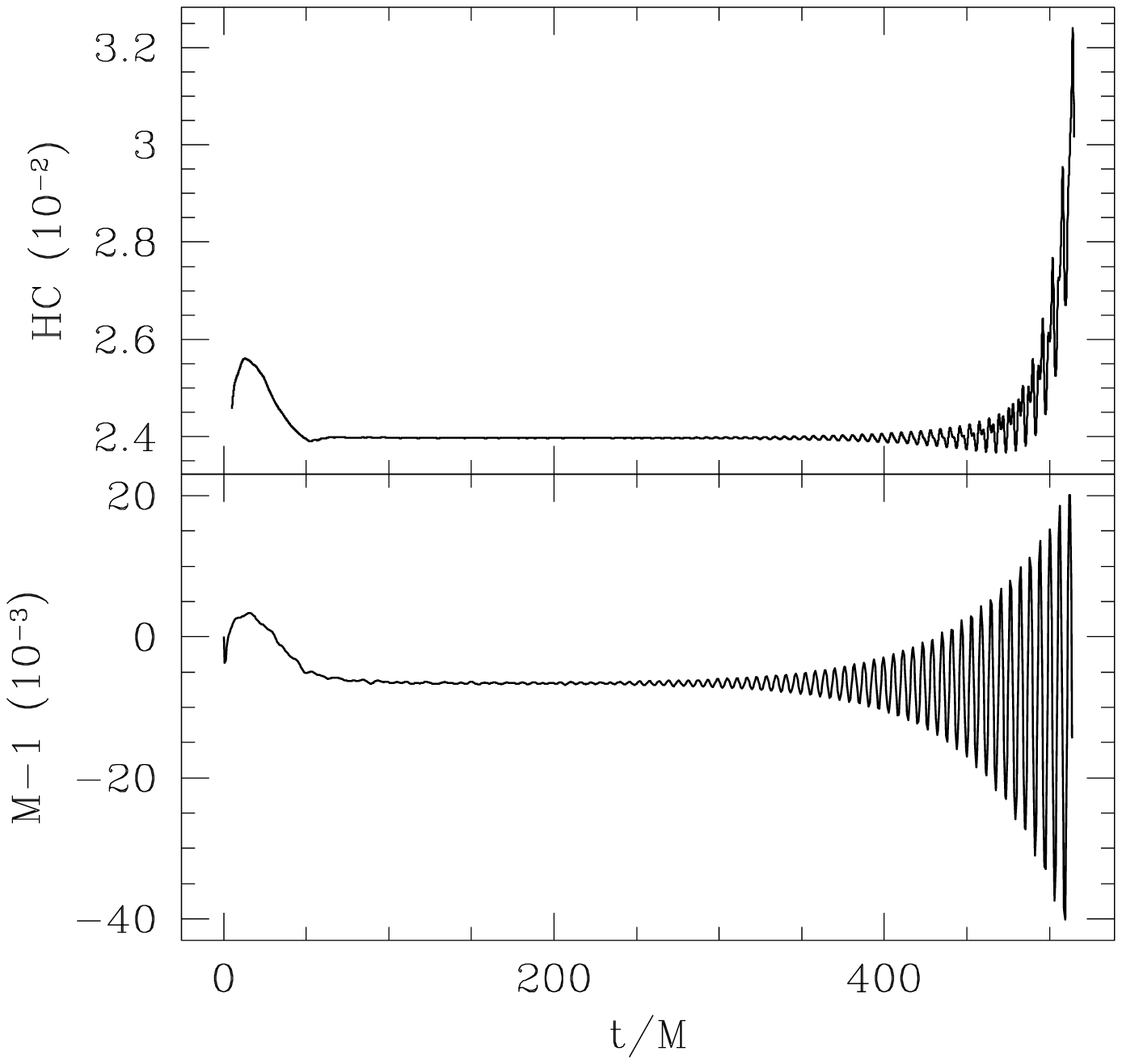}
\caption[figure1]{\label{fig2}
Same as in Fig.~\ref{fig1} but imposing reflection symmetry only across 
the $x=0$ plane.
}\end{figure*}

Figure~\ref{fig1} shows the L2 norm of the Hamiltonian constraint 
and the error in the mass 
of the black hole from a 3D evolution. 
The location of the outer boundary
of the computational domain was set to $9M$ 
and the grid-spacing to $0.25M$.  
We imposed quadrant symmetry, namely reflection 
symmetry across the $x=0$ and $y=0$ planes. 
This choice was motivated by the work in Ref.~\cite{miguel}, 
in which long term stability was only achieved in 
octant symmetry and with live gauge conditions. 
We performed evolutions with higher resolutions in octant symmetry only
for the purpose of testing second order convergence 
in our code. 
An identical, quadrant-symmetric,
run using the BSSN system crashed at $\sim 75M$.
Short-lived, BSSN evolutions with analytic lapse and shift were also
reported in Ref.~\cite{miguel}.
As one can see from Fig~\ref{fig1}, the new system is able to
extend the life-time of the simulation an order of magnitude beyond
that of BSSN.
To test the effects of the symmetry conditions, we repeated a simulation
using bitant or equatorial symmetry. Figure~\ref{fig2} shows the outcome of the
simulation. A comparison of the results in 
Figs.~\ref{fig1} and \ref{fig2}
indicate that the duration and quality of the data still
depend on the symmetries imposed.

Although we have managed to extend the duration of the simulation, as seen in
Fig.~\ref{fig2}, there is a late-onset of an instability.
Preliminary experiments indicate that the instability is likely to be
connected to the type of outer boundary condition and its placement,
as well as the symmetries imposed.
Our results are, nonetheless, comparable to evolutions in Ref.~\cite{KST} based on 
a parametrized hyperbolic formulation. In that study, 
evolutions lasting $\sim 600M$ were obtained for similar computational 
domain sizes, but without any symmetries applied, 
using analytic densitized lapse and shift conditions.  
A direct comparison with the results in \cite{KST} is not straightforward due to
the numerical technique employed, namely spectral methods.

Finally, we performed evolutions in octant symmetry for $1<n\le 2$. 
We obtained similar stability to that with $n=1$. 
The trend observed in these simulations suggests that for values $n>2$
it is very likely to experience a deterioration on the stability of the 
evolutions. This should not be surprising. Looking at the
conformal transformations (\ref{nct:K})-(\ref{nct:lapse}), one sees that
large values of $n$ have the
effect of increasing the spatial gradients of these primary variables.

\section{Conclusions}

This paper introduces a conformal-traceless 
formulation of the Einstein equation that yields a substantial
improvement in the duration of evolutions. 
This system has various elements in common with the BSSN system. However,
it has the following new aspects:
(1) modification of the conformal scalings of the extrinsic curvature,
(2) use of the mixed-index form of the traceless extrinsic curvature
as a primary variable, and
(3) densitization of the lapse function.
The reason behind these modifications is to eliminate
or switch the sign of non-linear terms involving the extrinsic curvature.
Our results support the view that these terms are 
partially responsible for rendering the evolutions unstable. 
We are currently investigating the degree to which each of the above
elements of the formulation here introduced
impacts the stability of the simulations\cite{ALS}.
We have demonstrated that our 3D simulations of a single black hole
exhibit a remarkable improvement when compared
with those using the BSSN formulation.  This is 
especially apparent when one considers we have performed these tests 
in what may be the most 
troublesome gauge choice, namely
obtained directly from the analytic black hole solution. 
Furthermore, we did not impose outgoing boundary conditions.
These conditions
have been shown to improve the duration of simulations in hyperbolic 
and BSSN systems.
We are currently extending our study to investigate whether 
the stability of this new system 
is preserved and improved using live gauge conditions 
and in dynamical situations 
such as distorted and binary black holes.

\section{Acknowledgments}
This work was supported by NSF grants
PHY-9800973 and PHY-0114375.
We want to thank A.~Ashtekar for comments and helpful discussions.
We also want to thank the rest the MAYA Project development team:
D.~Fiske, B.~Kelly, E.~Schnetter and K.~Smith.

\begin{thebibliography}{99}

\bibitem{ADM}
R. Arnowitt, S. Deser and C.W. Misner, in {\it Gravitation:
An Introduction to Current Research}, edited by L. Witten
(John Wiley, New York, 1962), pp. 227-265.

\bibitem{hyper_rev}
For reviews on hyperbolic methods
for Einstein's equations: ``The Cauchy
problem for the Einstein equations'', H. Friedrich and A. Rendall, in
{\it Einstein's Field Equations and their Physical Interpretation}, ed. by B. G.
Schmidt, Springer, Berlin, 2000.  Published in
{\it Lect. Notes Phys.} {\bf 540}, 127 (2000).
 e-Print Archive: gr-qc/000207. Also: ``Hyperbolic methods for Einstein's
equations'', O. Reula, {\it Living Reviews in Relativity}
(http://www.livingreviews.org)

\bibitem{KST}
Lawrence E. Kidder, Mark A. Scheel, Saul A. Teukolsky,
{\it Phys. Rev. D}, {\bf 64} 064017 (2001)

\bibitem{BS}
 T. Baumgarte and S. Shapiro,
{\it Phys. Rev. D}, {\bf 59} 024007 (1999).

\bibitem{SN}
 M. Shibata and T. Nakamura,
{\it Phys. Rev. D}, {\bf 52}, 5428 (1995).

\bibitem{miguel}
M. Alcubierre and B. Bruegmann, 
{\it Phys. Rev. D}, {\bf 63} 104006 (2001).

\bibitem{Kelly}
B. Kelly, P. Laguna, K. Lockitch, J. Pullin, 
E. Schnetter, D. Shoemaker and M. Tiglio
{\it Phys. Rev. D}, {\bf 64} 084013 (2001).

\bibitem{Kconst}
A.P. Gentle, D.E. Holz, A. Kheyfets, P. Laguna, W.A. Miller, D.M. Shoemaker
{\it Phys. Rev. D}, {\bf 63} 064024 (2001)

\bibitem{KLM}
H. Kurki-Suonio, P. Laguna and R.A. Matzner
{\it Phys. Rev. D}, {\bf 48}, 3611 (1993)

\bibitem{Luis}
L. Lehner, private communication (2001).

\bibitem{David}
D. Neilsen, in preparation (2002).

\bibitem{Centrella}
J. Centrella and J. Wilson, {\it Ap.J.}, {\bf 273}, 428 (1983).

\bibitem{Bardeen}
J.M. Bardeen and L.T. Buchman, gr-qc/0111085.

\bibitem{ALS}
A. Ashtekar, P. Laguna and D. Shoemaker, in preparation (2002).

\end{thebibliography}
\end{document}